\documentclass[aps,showpacs,superscriptaddress,nofootinbib,floatfix]{revtex4-1}
\usepackage{graphicx}
\usepackage{epsf}

\newcommand{\be}{\begin{equation}}
\newcommand{\ee}{\end{equation}}
\newcommand{\bea}{\begin{eqnarray}}
\newcommand{\eea}{\end{eqnarray}}
\newcommand{\beas}{\begin{eqnarray*}}
\newcommand{\eeas}{\end{eqnarray*}}
\newcommand{\bd}{\begin{displaymath}}
\newcommand{\ed}{\end{displaymath}}
\def\shiftdow_Kn#1{#1\llap{\low_Ker.04ex\hbox{#1}}}

\begin{document}
\title{$B\to \rho$ semileptonic decays and $|V_{ub}|$} 
\author{C.  \surname{Albertus}} 
\affiliation{Departamento
de F\'\i sica At\'omica, Nuclear y Molecular\\  e Instituto Carlos
 I de F\'{\i}sica Te\'orica y Computacional\\ Universidad de Granada, Avenida
 de Fuentenueva s/n, E-18071
Granada, Spain.} 
\author{E.  \surname{Hern\'andez}} 
\affiliation{Departamento
de F\'\i sica Fundamental e IUFFyM,\\ Universidad de Salamanca, Plaza de la
Merced s/n, E-37008
Salamanca, Spain.}  
\author{J.  \surname{Nieves}}
\affiliation{Instituto de F\'\i sica Corpuscular (IFIC), Centro Mixto
  CSIC-Universidad de Valencia, Institutos de Investigaci\'on de
  Paterna, Apartado 22085, E-46071 Valencia, Spain} 
  

\today

\begin{abstract}
We reevaluate the $B\to\rho\,l^+\nu_l$ decay width as a full $B\to\pi\pi\,l^+\nu_l$
four-particle decay, 
 in which 
the two final pions are produced via an intermediate $\rho$ meson. The decay width can be written
as a convolution of the $B\to\rho\,l^+\nu_l$ decay width, for
an off-shell $\rho$, with the $\rho\to\pi\pi$  line shape.
This allows to  fully incorporate the effects of the finite $\rho$
 meson width. As shown, consideration of the $\rho$ meson 
width effects  increase the $|V_{ub}|$ value by some 8\%, rendering it in
better agreement with the determination based in the
  $B\to\pi$ decay. We take the $q^2$ dependence of the $B\to \rho$
semileptonic form factors  from a dispersive Omn\`es representation. 
The Omn\`es subtraction constants
and the overall normalization parameter $|V_{ub}|$ are fitted to  light cone sum rules and 
lattice QCD  theoretical form-factor calculations, in the low and high $q^2$ regions
respectively, together to the  CLEO, BaBar and Belle  experimental
partial branching fraction distributions. The extracted value from
this global fit is
$|V_{ub}|=(3.12\pm 0.13)\times 10^{-3}$, in agreement  with the average
$B\to \pi$ exclusive value $|V_{ub}|=(3.23\pm 0.31)\times 10^{-3}$
 quoted by the Particle Data Group. The extracted value increases to 
$|V_{ub}|=(3.51\pm 0.16)\times 10^{-3}$ if only the most recent 
Belle Collaboration data
is used. This latter value is in agreement with different theoretical
determinations based in the $B\to\pi$ semileptonic decay and the values obtained
by the CKMfitter and UTfit groups. In any case a clear tension with the 
$|V_{ub}|$ value extracted from inclusive
semileptonic $b\to u$ decays still persists.
\end{abstract}
\maketitle

\section{Introduction}
 A precise determination of $V_{ub}$ is essential to check the consistency 
of the Standard Model, especially the description of $CP$ violations. However,
$V_{ub}$ is still the least well known element of the 
Cabibbo-Kobayashi-Maskawa (CKM) matrix. At present, there
is a clear tension between the
$|V_{ub}|$ values extracted from the analysis of inclusive and exclusive 
decays.  Determinations based on inclusive semileptonic decays have 
their largest uncertainties coming from the error on the $b-$ quark mass,
 but their
values tend to be consistent. From these analyses, the  average value quoted by
the Particle Data Group (PDG) in its 2013 update~\cite{Beringer:1900zz} is
 $|V_{ub}| = (4.41 \pm
0.15^{+0.15}_{-0.17}) \times10^{-3}$.  The corresponding average value 
extracted from  exclusive determinations is  dominated by the 
$B\to\pi$ semileptonic  decay value $|V_{ub}| =
(3.23 \pm 0.31)\times10^{-3}$~\cite{Beringer:1900zz}. In this case the
 error is dominated
by form factor normalizations. Another problem which will be addressed here, 
is the existing tension  between the exclusive
determinations using the $B\to\rho$ and $B\to\pi$ semileptonic 
decays. From $B\to\rho$ decays lower values have been traditionally reported, thus  for
instance, BaBar presented a value of   $|V_{ub}|=(2.75\pm0.24)\times
10^{-3}$ in \cite{delAmoSanchez:2010af}, while in the  approach of
Ref.~\cite{Flynn:2008zr}, similar to
the one followed here and based on the Omn\`es representation of the
form factors, was obtained $|V_{ub}|=(2.76\pm0.21)\times
10^{-3}$. Very recent analyses,   using light cone sum rules 
 (LCSR), also find central values, $|V_{ub}|=(2.91\pm0.19)\times 10^{-3}$  and 
 $|V_{ub}|=(3.11\pm0.19)\times 10^{-3}$ ~\cite{Fu:2014pba}, below
those found from $B\to \pi$  decays ($|V_{ub}|=(3.47\pm0.29\pm0.03)
\times 10^{-3}$~\cite{Flynn:2007ii},
$|V_{ub}|=(3.6\pm0.4_{\rm stat}\pm0.2^{+0.6}_{{\rm syst}-0.4{\rm thy}})\times 10^{-3}$~\cite{Adam:2007pv},
$|V_{ub}|=(3.41^{+0.37}_{-0.32}|_{\rm th} \pm 0.06|_{\rm exp}) \times10^{-3}$
~\cite{Khodjamirian:2011ub}, $|V_{ub}|=(3.52\pm0.29)\times 10^{-3}$~\cite{Sibidanov:2013rkk}).  As
  pointed out in Ref.~\cite{Meissner:2013pba},  part of this
  systematic discrepancy
   could be due to the fact that the $B\to\rho$ analyses do not take
    into account
  the effect of the broad $\rho-$width. 
  
  In Ref.~\cite{Kang:2013jaa} the authors propose
  to extract $|V_{ub}|$ from the analysis of the four-body semileptonic decay
  $B\to\pi\pi l^+\nu_l$ taking into account $\pi\pi$ rescattering effects and
  the effect of the rho meson. Their approach is based on dispersion theory and
  does not rely on specific resonant contributions. In our calculation we 
  do a simpler study of the four-body decay in which 
  the two pions are produced via an intermediate  $\rho$ meson $B\to\pi\pi(\rho)l^+\nu_l$. 
  The decay width  can then be expressed as an integration over the $\rho$ meson invariant
   mass   available in the $B\to\rho\, l^+\nu_l$ decay 
  for an off-shell $\rho$, weighted by the $\rho\to\pi\pi$ line shape distribution 
  that fully takes into 
  account  $\rho$ meson width effects.  
 In fact, this type of analysis has been recently done  by the
 Belle collaboration  in Ref.~\cite{Sibidanov:2013rkk}
 with the result that a  larger $|V_{ub}|$ value, in better agreement with the determination
 from  $B\to\pi$ semileptonic decay, is obtained. 

 In this work we perform a combined fit
 to the latest partial branching fraction distributions  by the different experimental collaborations, while at the same time
 we substantially improve on the  treatment of the form factors over previous works. 
 In this respect we shall follow Ref.~\cite{Flynn:2008zr}, where the $B\to\rho$  form
  factors are described using a multiply subtracted Omn\`es dispersion
  relation. The Omn\`es functional form depends on the form factor values at
  the subtraction points and those values are treated as free parameters. These,
  together with  $|V_{ub}|$, are fitted both to $B\to\rho\, l^+\nu_l$
  recent  
  partial branching fraction  measurements from 
  Belle~\cite{Sibidanov:2013rkk},  BaBar~\cite{delAmoSanchez:2010af} and
  CLEO~\cite{Gray:2007pw}
  collaborations, as well as to theoretical results for the $B\to\rho$
   form factors obtained  using  LCSR~\cite{Ball:2004rg}
   and lattice calculations by the SPQcdR~\cite{Abada:2002ie} and 
   UKQCD~\cite{Bowler:2004zb} collaborations.
   For the $\rho\to\pi\pi$ decay we use a
  phenomenological vertex where the coupling constant has been fixed to the on-shell
  $\rho$ meson decay width.

 The paper is organized as follows. In Sec.\ref{sec:gamma}  we present all the 
 expressions  needed to evaluate the decay width. We shall give a triple 
 differential decay width distribution with respect to $p_\rho^2, q^2$ 
 and $x_l$, with $p_\rho^2$ the $\rho$ meson invariant mass square, $q$ the 
 total four-momentum
  of the final lepton system, and $x_l$ the cosinus of the angle formed by the
 momentum of the charged lepton, measured in the lepton center   of mass system,
 and 
 the momentum of the virtual $\rho$ in the $B$ meson rest  frame. These are 
 the variables used by the experiments, and in order to obtain the fractional 
 branching fractions (see below) we just have to integrate over their corresponding
  ranges. Sec.~\ref{sec:fit} 
 describes the fitting procedure that follows closely
 Ref.~\cite{Flynn:2008zr}, and finally, in Sec.~\ref{sec:res}
 we present and discuss the main results of this work.  
 In Appendix~\ref{app:ha}, we give details on the helicity amplitude
 formalism used to evaluate   the product of the
 leptonic and hadronic tensors, while  in Appendix~\ref{app:gcm} we provide the
 correlation matrix resulting from our global fit.
\vspace{-.0cm}\section{$\Gamma[B\to\pi\pi(\rho)\,l^+\nu_l]$ decay width}
\label{sec:gamma}
\begin{figure}[h!!]
\begin{center}
\vspace{1cm}\resizebox{8.cm}{!}{\includegraphics{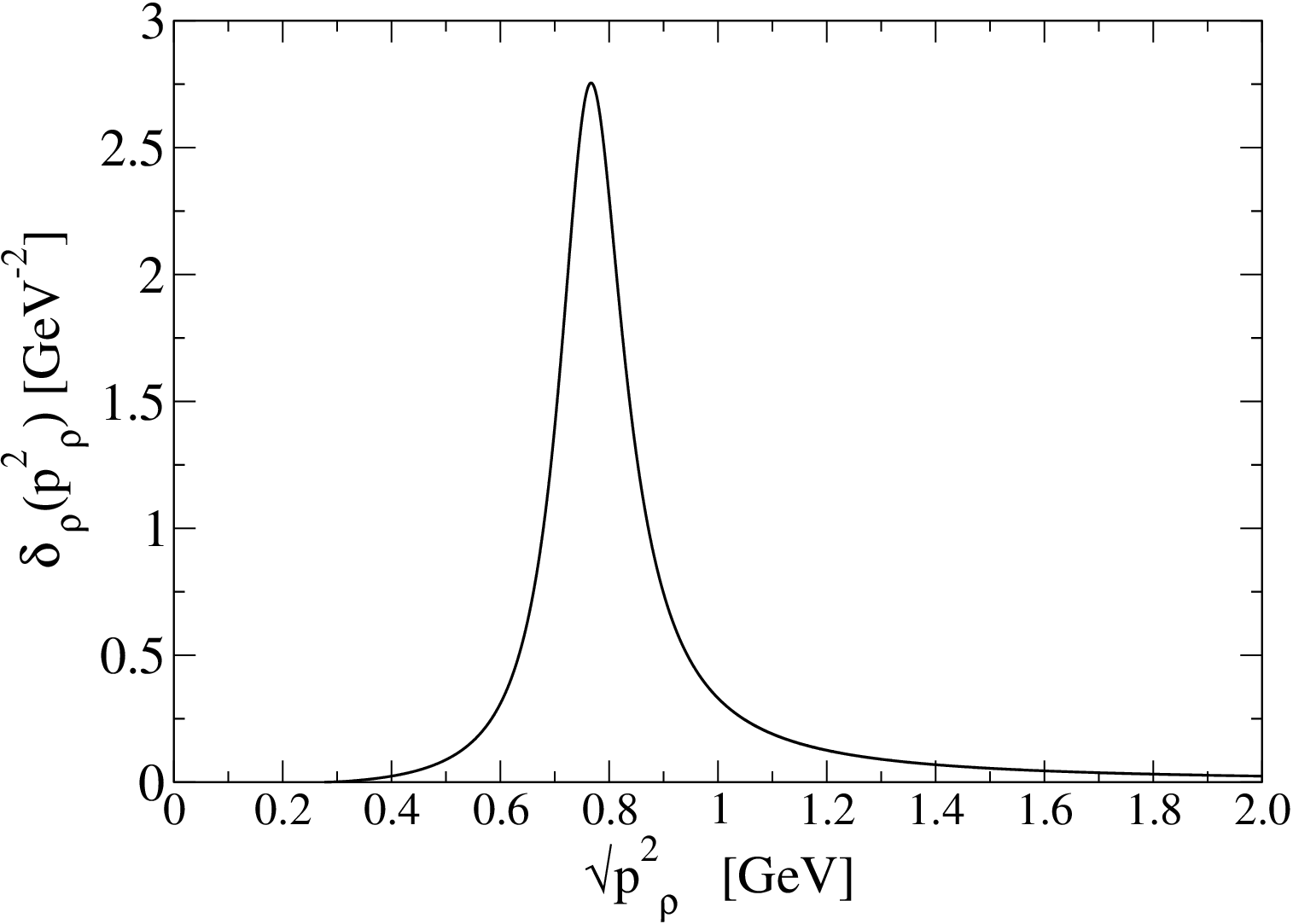}}
\caption{$\delta_\rho(p_\rho^2)$ representation as a function of
$\sqrt{p_\rho^2}$.}
\label{fig:deltarho}
\end{center}
\end{figure}

\vspace{-.15cm}
\noindent
Working in the exact isospin limit, the $B\to\pi\pi(\rho)\,l^+\nu_l$ decay width  is given by
\bea
&&\hspace{-1cm}\Gamma=\left(\frac{G_F}{\sqrt2}\right)^2|V_{ub}|^2C_\rho^2\frac1{2m_B}
\frac1{(2\pi)^{8}}\int\frac{d^3p_l}{2E_l}\int
\frac{d^3p_\nu}{2E_\nu}\int\frac{d^3p_{\pi_1}}{2E_{\pi_1}}\int
\frac{d^3p_{\pi_2}}{2E_{\pi_2}}
\,\delta^{(4)}(p_B-p_l-p_\nu-p_{\pi_1}-p_{\pi_2})\nonumber\\
&&\hspace{2cm}\times\sum_{s_l}\sum_{s_\nu}\Big|h_{\alpha\sigma}(p_B,p_\rho)
\frac{-g^{\sigma\beta}+
\frac{p_{\rho}^\sigma p_{\rho}^\beta}{p_\rho^2}}
{p_\rho^2-m_\rho^2+i\sqrt{p_\rho^2}\,\Gamma_\rho(p_\rho^2)}(p_{\pi_1}
-p_{\pi_2})_\beta\,
\bar u_{s_\nu}(p_\nu)\gamma^\alpha(1-\gamma_5)v_{s_l}(p_l)\Big|^2,
\eea
where only the transverse part of the $\rho$ meson propagator contributes
in that limit~\cite{LopezCastro:1999xg}. $G_F=1.166378\times 10^{-5}\,
$GeV$^{-2}$~\cite{Beringer:1900zz}  is
the Fermi decay constant and  $C_\rho= 6$ is the effective $\rho\to\pi\pi$ 
coupling constant  with
\bea
\Gamma_\rho(p_\rho^2)=\frac{C_\rho^2}{6\pi\,p_\rho^2}
\Big(\frac{p_\rho^2}{4}-m_\pi^2\Big)^{3/2}
\eea
 being the $\rho$ meson width for $\sqrt{p_\rho^2}$
invariant mass. Besides,
 $p_B=(m_B,\vec 0\,)$, $p_\rho=p_B-p_l-p_\nu$ and
\bea
h_{\alpha\sigma}(p_B,p_\rho)&=&\frac{2V(q^2)}{m_B+\sqrt{p_\rho^2}}\,
\epsilon_{\alpha\gamma\delta\sigma}p_B^\gamma p_\rho^\delta
-i(m_B+\sqrt{p_\rho^2}\,)A_1(q^2)\,g_{\alpha\sigma}\nonumber\\
&&+i
\frac{A_2(q^2)}{m_b+\sqrt{p_\rho^2}}\,q_{\sigma}(p_B+p_\rho)_\alpha\,
-i
\frac{2A(q^2)}{q^2}\sqrt{p_\rho^2}\ q_\alpha(p_B+p_\rho)_\sigma,
\eea
where $V$ and $A_1,A_2,A$ are respectively the vector and axial form factors for
the $B\to\rho$ weak transition. Here we use $\epsilon_{0123}=+1$ and we have
defined  $q=p_B-p_\rho=p_l+p_\nu$, which is the total four-momentum carried by 
the leptons. In the above expression for $h_{\alpha\sigma}$ we have substituted
$m_\rho$ by $\sqrt{p_\rho^2}$ with respect to the corresponding expression in 
Ref.~\cite{Flynn:2008zr}\\
\noindent
The above expression for $\Gamma$ can be rewritten as
\bea
\Gamma&=&
\left(\frac{G_F}{\sqrt2}\right)^2|V_{ub}|^2\frac1{m_B}
\frac1{(2\pi)^{6}}\int\frac{d^3p_l}{E_l}\int
\frac{d^3p_\nu}{E_\nu}\ {\cal L}^{\alpha\alpha'}(p_l,p_\nu)\nonumber\\
&&\hspace{2cm}\times\sum_{r=\pm1,0}\sum_{s=\pm1,0}\frac{h_{\alpha\sigma}(p_B,p_\rho)\epsilon_r^{\sigma\,*}(p_\rho)
h^*_{\alpha'\sigma'}(p_B,p_\rho)\epsilon_s^{\sigma'\,}(p_\rho)}
{|p_\rho^2-m_\rho^2+i\sqrt{p_\rho^2}\,\Gamma_\rho(p_\rho^2)|^2}\
2\sqrt{p_\rho^2}\,\Gamma_{\rho}^{rs}(p_\rho^2)
\eea
where we have used that
\bea
\big(-g^{\sigma\beta}+
\frac{p_{\rho}^\sigma p_{\rho}^\beta}{p_\rho^2}\Big)
=\sum_{r=\pm1,0}\epsilon_r^{\sigma\,*}(p_\rho)
\epsilon_r^{\beta}(p_\rho),
\eea
with $\epsilon_r(p_\rho),\, r=\pm1,0$ the three polarization vectors of a 
$\rho$ meson with invariant mass given by
$\sqrt{p_\rho^2}$.  ${\cal L}^{\alpha\alpha'}(p_l,p_\nu)$ is
 the lepton tensor given by
\bea
{\cal L}^{\alpha\alpha'}(p_l,p_\nu)=p_l^\alpha p_\nu^{\alpha'}-g^{\alpha\alpha'}p_l\cdot
p_\nu+p_l^{\alpha'}
p_\nu^{\alpha}\pm i\epsilon^{\gamma\alpha\delta\alpha'}p_{l\gamma}p_{\nu\,\delta},
\eea
 where the $\pm$ sign corresponds to $l^+\nu_l$ or $l\,\bar\nu_l$ decays respectively
and
\bea
\Gamma_{\rho}^{rs}(p_\rho^2)=C_\rho^2\frac{\epsilon_{r}^\beta(p_\rho)
\epsilon_s^{\beta'\,*}(p_\rho)}{2\sqrt{p_\rho^2}\,(2\pi)^2}\int\frac{d^3p_{\pi_1}}{2E_{\pi_1}}\int\frac{d^3p_{\pi_2}}{2E_{\pi_2}}
\delta^{(4)}(p_\rho-p_{\pi_1}-p_{\pi_2})
(p_{\pi_1}-p_{\pi_2})_\beta\,
(p_{\pi_1}-p_{\pi_2})_{\beta'}.
\label{eq:grho}
\eea
The integrals in $\Gamma_{\rho}^{rs}(p_\rho^2)$ can be
readily evaluated using Lorentz covariance and one gets that
\bea
\Gamma_{\rho}^{rs}(p_\rho^2)=-\delta_{rs}\,
\Gamma_\rho(p_\rho^2),
\eea
Then,
\bea
&&\hspace{-1cm}\Gamma=\left(\frac{G_F}{\sqrt2}\right)^2|V_{ub}|^2\frac1{m_B}
\frac1{(2\pi)^{5}}\int\frac{d^3p_l}{E_l}\int
\frac{d^3p_\nu}{E_\nu}{\cal L}^{\alpha\alpha'}(p_l,p_\nu)\ {\cal
H}_{\alpha\alpha'}(p_B,p_\rho)\, \delta_{\rho}(p_\rho^2),
\eea
where we have defined the hadronic tensor
\bea
{\cal
H}_{\alpha\alpha'}(p_B,p_\rho)=\sum_{r=\pm1,0}h_{\alpha\sigma}(p_B,p_\rho)\epsilon_r^{\sigma\,*}(p_\rho)
h^*_{\alpha'\sigma'}(p_B,p_\rho)\epsilon_r^{\sigma'\,}(p_\rho),
\eea
and the $\rho$ meson line shape function
\bea
\delta_{\rho}(p_\rho^2)=\frac1\pi\frac{\sqrt{p_\rho^2}\,\Gamma_{\rho}(p_\rho^2)}
{|p_\rho^2-m_\rho^2+i\sqrt{p_\rho^2}\,\Gamma_\rho(p_\rho^2)|^2}.
\eea
A representation of the latter as a function of the $\rho$ invariant mass
 is given in
Fig.~\ref{fig:deltarho}.
In the  $\Gamma_\rho(p_\rho^2)\to 0$ limit, one would have
\bea
\delta_\rho(p_\rho^2)\approx 
\delta(p_\rho^2-m_\rho^2),
\eea
and in that case $\Gamma$ would be given
by
\bea
\label{eq:rhosinanchura}
\Gamma&\approx&\left(\frac{G_F}{\sqrt2}\right)^2|V_{ub}|^2\frac1{m_B}
\frac1{(2\pi)^{5}}\int\frac{d^3p_l}{E_l}\int
\frac{d^3p_\nu}{E_\nu}{\cal L}^{\alpha\alpha'}(p_l,p_\nu)
{\cal
H}_{\alpha\alpha'}(p_B,p_\rho)
\,\delta({p_\rho^2}-m^2_\rho)\nonumber\\
%
%
%
&=&\left(\frac{G_F}{\sqrt2}\right)^2|V_{ub}|^2\frac1{m_B}
\frac1{(2\pi)^{5}}\int\frac{d^3p_l}{E_l}\int\frac{d^3p_\nu}{E_\nu}
\int{d^4p_\rho}
\,\delta(p_\rho^2-m_\rho^2)
\delta^{(4)}(p_B-p_l-p_\nu-p_\rho)\,{\cal L}^{\alpha\alpha'}(p_l,p_\nu)
{\cal
H}_{\alpha\alpha'}(p_B,p_\rho)\nonumber\\
&=&\left(\frac{G_F}{\sqrt2}\right)^2|V_{ub}|^2\frac1{m_B}
\frac1{(2\pi)^{5}}\int\frac{d^3p_l}{E_l}\int\frac{d^3p_\nu}{E_\nu}
\int\frac{d^3p_\rho}{2E_\rho}
\delta^{(4)}(p_B-p_l-p_\nu-p_\rho)\,{\cal L}^{\alpha\alpha'}(p_l,p_\nu)
{\cal
H}_{\alpha\alpha'}(p_B,p_\rho)\nonumber\\
&=&\Gamma(B\to\rho\, l^+\nu_l)
\eea
recovering the expression for the $B\to\rho\, l^+\nu_l$ decay width used in
previous analyses where the $\rho$ meson widths effects were not taken into
account.

Going back to the full expression, it can be  rewritten  as
\beas
\Gamma&=&\left(\frac{G_F}{\sqrt2}\right)^2\frac{|V_{ub}|^2}{m_B(2\pi)^{5}}
\int\frac{d^3p_\nu}{E_\nu}\int
\frac{d^3p_l}{E_l}{\cal L}^{\alpha\alpha'}(p_l, p_\nu)\,
{\cal H}_{\alpha\alpha'}(p_B,p_B-p_l-p_\nu )\, \delta_\rho(p_\rho^2)
\\
&=&\left(\frac{G_F}{\sqrt2}\right)^2\frac{|V_{ub}|^2}{m_B(2\pi)^{5}}
\int dp_\rho^2\, \delta_\rho(p_\rho^2)\,\int\frac{d^3p_\rho}{2E_\rho}
\,\int\frac{d^3p_\nu}{E_\nu}\int
\frac{d^3p_l}{E_l}{\cal L}^{\alpha\alpha'}(p_l, p_\nu)\,
{\cal H}_{\alpha\alpha'}(p_B,p_\rho )\delta^{(4)}(p_B-p_\rho-p_l-p_\nu)
\eeas
with 
$E_\rho=p_\rho^0=\sqrt{p_\rho^2+\vec p_\rho^{\ ^2}}$. Then,
\bea
\Gamma&=&\int dp_\rho^2\, \delta_\rho(p_\rho^2)\,
\Big\{\left(\frac{G_F}{\sqrt2}\right)^2
\frac{|V_{ub}|^2}{m_B(2\pi)^{5}}
\int \frac{d^3p_\rho}{2E_\rho}\int\frac{d^3p_\nu}{E_\nu}\int
\frac{d^3p_l}{E_l}{\cal L}^{\alpha\alpha'}(p_l, p_\nu)\,
{\cal H}_{\alpha\alpha'}(p_B,p_\rho
)\delta^{(4)}(p_B-p_\rho-p_l-p_\nu)\Big\},\nonumber\\
\eea
where the term in curly brackets represents the $B\to\rho\, l^+\nu_l$ decay width 
for the case of a final $\rho$ meson with invariant mass 
$\sqrt{p_\rho^2}$. The integrals on neutrino variables can be evaluated using Lorentz covariance 
\bea
\int\frac{d^3p_\nu}{E_\nu}\int
\frac{d^3p_l}{E_l}{\cal L}^{\alpha\alpha'}(p_l, p_\nu)\,
{\cal H}_{\alpha\alpha'}(p_B,p_\rho )\delta^{(4)}(q-p_l-p_\nu)
=2\pi\frac{q^2-m_l^2}{2q^2}\int_{-1}^1 dx_l\,{\cal L}^{\alpha\alpha'}(\tilde p_l, 
\tilde p_\nu)
{\cal H}_{\alpha\alpha'}(\Lambda p_B,\Lambda p_\rho ),
\eea
where  $\Lambda$ is a rotation that takes $\vec p_\rho$ to the
negative $Z$ axis followed by a boost to the center of mass of the two final
leptons. In that case
\bea
\Lambda
p_B&=&\frac1{2\sqrt{q^2}}\,\Big(m_B^2+q^2-p_\rho^2,0,0,
-\lambda^{1/2}(m_B^2,q^2,p_\rho^2)   \Big),\nonumber\\ 
\Lambda p_\rho&=&\frac1{2\sqrt{q^2}}\,\Big({m_B^2-q^2-p_\rho^2},0,0,
-\lambda^{1/2}(m_B^2,q^2,p_\rho^2)\Big).
\eea
It is clear now that the product of tensors ${\cal L}^{\alpha\alpha'}(\tilde p_l, 
\tilde p_\nu)
{\cal H}_{\alpha\alpha'}(\Lambda p_B,\Lambda p_\rho )$ does not depend on 
the lepton 
$\varphi_l$ azimuthal angle that
can then be integrated out to give a factor $2\pi$. The lepton $\tilde p_l$ and
$\tilde p_\nu$ momenta can be chosen for simplicity as
\bea
\tilde p_l=(\frac{q^2+m_l^2}{2\sqrt{q^2}},\frac{q^2-m_l^2}{2\sqrt{q^2}}\sqrt{1-x_l^2},0,
-\frac{q^2-m_l^2}{2\sqrt{q^2}}x_l),\\
\tilde p_\nu=(\frac{q^2-m_l^2}{2\sqrt{q^2}},-\frac{q^2-m_l^2}{2\sqrt{q^2}}\sqrt{1-x_l^2},0,
\frac{q^2-m_l^2}{2\sqrt{q^2}}x_l).
\eea
With this definition, 
$x_l$ is the cosinus of the angle formed by the momentum of the charged lepton 
measured in 
the center of mass of the two leptons, with the direction of 
the momentum of the virtual $\rho$ meson measured in the reference frame in 
which the $B$ meson is at rest. Since there is no dependence on the $\vec p_\rho$
angular variables we find
\bea
\Gamma&=&\int dp_\rho^2\, \delta_\rho(p_\rho^2)\,
\Big\{\left(\frac{G_F}{\sqrt2}\right)^2
\frac{|V_{ub}|^2}{2m_B(2\pi)^{3}}\int\frac{\lambda^{1/2}(m_B^2,q^2,p_\rho^2)}{2m_B}dE_\rho
\frac{q^2-m_l^2}{2q^2}\int_{-1}^1 dx_l\,{\cal L}^{\alpha\alpha'}(\tilde p_l, 
\tilde p_\nu)
{\cal H}_{\alpha\alpha'}(\Lambda p_B,\Lambda p_\rho )\Big\},
\eea
from where one can  write the
following differential decay width
\bea
\frac{d\Gamma}{dp_\rho^2\,dq^2\,dx_l}=
 \delta_\rho(p_\rho^2)\,\frac{G_F^2|V_{ub}|^2}{128\pi^3m_B^3q^2}
\lambda^{1/2}(m_B^2,q^2,p_\rho^2)
({q^2-m_l^2})\,{\cal L}^{\alpha\alpha'}( \tilde p_l, \tilde p_\nu)
\,{\cal H}_{\alpha\alpha'}(\Lambda p_B,
\Lambda p_\rho ).
\eea
The product ${\cal L}^{\alpha\alpha'}( \tilde p_l, \tilde p_\nu)
\,{\cal H}_{\alpha\alpha'}(\Lambda p_B,
\Lambda p_\rho )$ can be evaluated using the formalism of helicity amplitudes 
(see for instance Ref.~\cite{Ivanov:2005fd}) that we discuss in Appendix~\ref{app:ha}.
The final expression for the triple differential decay width is
\bea
\frac{d\Gamma}{dp_\rho^2\,dq^2\,dx_l}&=&
 \delta_\rho(p_\rho^2)\,\frac{G_F^2|V_{ub}|^2}{512\pi^3m_B^3q^2}
\lambda^{1/2}(m_B^2,q^2,p_\rho^2)
({q^2-m_l^2})^2\nonumber\\
&&\hspace{1cm}\times\Big\{\
2(1-x_l^2)H_{00}+(1\mp x_l)^2H_{+1+1}+(1\pm x_l)^2H_{-1-1}\nonumber\\
&&\hspace{1.5cm}
+\frac{m_l^2}{q^2}[(1-x_l^2)(H_{+1+1}+H_{-1-1})+2x_l^2H_{00}+2H_{tt}+
4x_lH_{t0}]\Big\}.
\eea
where the different $H_{rs}$ hadronic helicity amplitudes are defined and given in 
Appendix~\ref{app:ha}. Besides de upper sign corresponds to $l^+\nu_l$ decays, 
like experiments in Refs.~\cite{Gray:2007pw,delAmoSanchez:2010af,Hokuue:2006nr}, 
 while
the lower sign corresponds to $l^-\bar\nu_l$ ones, like in the latest Belle~\cite{Sibidanov:2013rkk}
 analysis. This difference is only relevant if the integration over $x_l$ does not cover
 its full range $[-1,1]$ as in the case of CLEO data~\cite{Gray:2007pw}.
Neglecting lepton masses, a good approximation for light $l=e,\mu$ final
 leptons,
 one arrives at the  expression
\bea
\frac{d\Gamma}{dp_\rho^2\,dq^2\,dx_l}\approx
\delta_\rho(p_\rho^2)\,\frac{G_F^2|V_{ub}|^2}{512\pi^3m_B^3}q^2
\lambda^{1/2}(q^2,m_B^2,p_\rho^2)
\Big[ 2(1-x_l)^2\,{\cal H}_{00}+(1\mp x_l)^2
\,{\cal H}_{+1+1}+(1\pm x_l)^2\,{\cal H}_{-1-1}\Big].
\eea
where the corresponding helicity amplitudes depend only on
the $V(q^2),A_1(q^2)$ and $A_2(q^2)$ form factors. 
%
%
%
%
\section{Fitting procedure}
\label{sec:fit}
The fitting procedure that we shall use  is, with minor modifications, 
the one followed in Ref.~\cite{Flynn:2008zr}. We describe our $B\to\rho$
form factors using a multiply subtracted Omn\`es dispersion 
relation~\cite{omnes1,omnes2}, the latter being based in unitarity and analyticity.
We will have
\bea
F(q^2)=\frac1{s_0-q^2}\prod_{j=0}^n[F(q_j^2)(s_0-q_j^2)]^{\alpha_j(q^2)}\ \ ,\ \
 \alpha_j(q^2)=\prod_{\stackrel{k=0}{k\neq j}}^n\frac{q^2-q_k^2}{q_j^2-q_k^2},
 \ \ F=V,A_1,A_2.
\eea
$s_0$ corresponds to the pole of the form factor and we shall use 
$s_0=m_{B^*}=5.3252\,$GeV~\cite{Beringer:1900zz} for the vector form factor 
and $s_0=5.7235$\,GeV~\cite{Beringer:1900zz} (the
mass of the $1^+$ B meson) for the two axial form factors.
As in Ref.~\cite{Flynn:2008zr}, we use three subtraction points at $q^2=0,\,
2q^2_{\rm max}/3,\,q^2_{\rm max}$ where we take   
$q^2_{max}=(m_B-m_\rho)^2=20.3\,$GeV$^2$ as used in 
Refs.~\cite{Gray:2007pw,delAmoSanchez:2010af,Hokuue:2006nr}. Note however the
latest Belle analysis in Ref.~\cite{Sibidanov:2013rkk} works with $q^2$ values up to
22-24\,GeV$^2$. The values of
$F(0),F(2q^2_{\rm max}/3)$ and $F(q^2_{\rm max})$, for $F=V,A_1,A_2$, are treated
as free parameters as will be  $|V_{ub}|$. The values of these ten
parameters are then fitted to reproduce form factor theoretical results obtained 
in LCSR~\cite{Ball:2004rg} and lattice 
calculations~\cite{Abada:2002ie,Bowler:2004zb}, and experimental 
measurements of partial branching fractions
obtained by the  CLEO~\cite{Gray:2007pw},  BaBar~\cite{delAmoSanchez:2010af} and 
 Belle~\cite{Sibidanov:2013rkk} 
collaborations. The partial branching fractions are defined as
\bea
B=\frac{1}{\Gamma(B^0)}\int_{p_{\rho\,
{\rm inf}}^2}^{p_{\rho\, {\rm sup}}^2}dp_\rho^2
\int_{q^2_{\rm inf}}^{q^2_{\rm sup}}dq^2\int_{x_{l\, {\rm inf}}}
^{x_{l\, {\rm sup}}}dx_{l}\,\frac{d\Gamma}
{dp_\rho^2\,dq^2\,dx_{l}},
\label{eq:pbf}
\eea
where for  the $B^0$ lifetime we use 
$\tau_0=1/\Gamma(B^0)=(1.519\pm0.007)\times 
10^{-12}\,$s~\cite{Beringer:1900zz}. It is worth mentioning  that even
  though the different experiments select $\rho$ events in a reduced
  $\sqrt{p_\rho^2}$ interval\footnote{ Both CLEO~\cite{Gray:2007pw} and 
  Belle~\cite{Sibidanov:2013rkk} accept $\rho$ events for $\sqrt{p_\rho^2}$
in the interval $m_\rho\pm2\Gamma_\rho$ while for 
BaBar~\cite{delAmoSanchez:2010af} the corresponding interval is
$[0.65,0.85]$\,GeV.}, this is treated as an overall acceptance effect
   that it is corrected in the final data~\cite{bernlochner}. Thus,  one has that
   $p_{\rho\, {\rm sup}}^2=(m_B-m_l)^2$ and $p_{\rho\, {\rm inf}}^2=4m_\pi^2$.
   For the lower (inf) and upper (sup)
 limits in $q^2$ and $x_l$  we use the values provided by the
  experiments (see Table~\ref{tab:ed}).
    For the $B^+$ lifetime to be used below
  we take $\tau_+=(1.641\pm0.008)\times 
10^{-12}\,$s~\cite{Beringer:1900zz}.
%
%
%
\subsection{Experimental and theoretical input}
Experimental data by the  CLEO~\cite{Gray:2007pw}, 
 BaBar~\cite{delAmoSanchez:2010af} and Belle~\cite{Sibidanov:2013rkk} 
 collaborations consist of partial
branching fractions as defined in Eq.(\ref{eq:pbf}). Their values together
with statistical and systematic errors  are collected in 
Table~\ref{tab:ed}. CLEO  has made used of isospin symmetry
to combine results for neutral and charged $B$ meson
decays. For BaBar data we have combined their $B^0\to\rho^-l^+\nu_l$ 4-mode and 
$B^+\to\rho^0 l^+\nu_l$ data  in the
following way:  
Denoting as $\sigma$ and $\epsilon$ the statistical
and systematic errors respectively we have evaluated
\bea
\frac1{\sigma^2}&=&\frac1{\sigma^2_{\rho^-}}+\frac1{(2\frac{\tau_0}{\tau_+}\,\sigma_{\rho^0})^2},\nonumber\\
\frac1{\epsilon^2}&=&\frac1{\epsilon^2_{\rho^-}}+
\frac1{(2\frac{\tau_0}{\tau_+}\,\epsilon_{\rho^0})^2},\nonumber\\
\frac{ B}{\sigma^2+\epsilon^2}&=&
\frac{ B_{\rho^-}}{\sigma^2_{\rho^-}+\epsilon^2_{\rho^-}}
+\frac{2\frac{\tau_0}{\tau_+}\, B_{\rho^0}}{(2\frac{\tau_0}{\tau_+}\,\sigma_{\rho^0})^2+(2\frac{\tau_0}{\tau_+}\,\epsilon_{\rho^0})^2}
\eea
In the case o the newest Belle's data~\cite{Sibidanov:2013rkk} we treat separately the neutral and
charge meson decays since they have been evaluated for different $q^2$ bins. However in order to
perform the fit we multiply the $\rho^0$ data by $2\tau_0/\tau_+$. 
\begin{table}
\begin{tabular}{lccc}\hline\hline
&$q^2\,[{\rm GeV^2}]$&$x_{l}$  &$10^4B$\\\hline
{\rm CLEO}~\cite{Gray:2007pw}\hspace{1cm}
&$0-2$&\hspace*{.15cm}$[-1,1]$\hspace*{.15cm}&$0.45\pm0.20\pm0.15$\\
&$2-8$&$[-1,1]$&$0.96\pm0.20\pm0.29$\\
&$8-16$&$[0,1]$&$0.75\pm0.16\pm0.14$\\
&$16-20.3$&$[0,1]$&$0.35\pm0.07\pm0.05$\\
&$8-20.3$&$[-1,0]$&$0.42\pm0.18\pm0.31$\vspace{.25cm}\\
{\rm BaBar}~\cite{delAmoSanchez:2010af}
&$0-8$&$[-1,1]$    &$0.587\pm0.084\pm0.097$\\
&$8-16$&$[-1,1]$   &$0.928\pm0.047\pm0.103$\\
&$16-20.3$&$[-1,1]$&$0.263\pm0.017\pm0.042$\vspace{.25cm}\\
{\rm Belle}~\cite{Sibidanov:2013rkk}\\
 $\rho^+$ data
&$0-4$&$[-1,1]$  &$0.373\pm0.106$\\
&$4-8$&$[-1,1]$  &$0.718\pm0.116$\\
&$8-12$&$[-1,1]$ &$0.806\pm0.123$\\
&$12-16$&$[-1,1]$&$0.723\pm0.125$\\
&$16-20$&$[-1,1]$&$0.626\pm0.115$\\
&$20-24$&$[-1,1]$&$0.017\pm0.079$\vspace{.25cm}\\
 $\rho^0$ data $\times\ 2\tau_0/\tau_+$\hspace{.5cm}
&$0-2$&$[-1,1]$ &$0.2296\pm0.0629$\\
&$2-4$&$[-1,1]$ &$0.2851\pm0.0574$\\
&$4-6$&$[-1,1]$ &$0.3314\pm0.0629$\\
&$6-8$&$[-1,1]$ &$0.4017\pm0.0629$\\
&$8-10$&$[-1,1]$&$0.2647\pm0.0537$\\
&$10-12$&$[-1,1]$&$0.3684\pm0.0629$\\
&$12-14$&$[-1,1]$&$0.4147\pm0.0629$\\
&$14-16$&$[-1,1]$&$0.4017\pm0.0611$\\
&$16-18$&$[-1,1]$&$0.3240\pm0.0592$\\
&$18-20$&$[-1,1]$&$0.2647\pm0.0180$\\
&$20-22$&$[-1,1]$&$0.1092\pm0.0481$\\
\hline\hline
\end{tabular}
\caption{Experimental partial branching fractions used as input. The different
$q^2$ and $x_l$ intervals are shown. Belle's original 
$\rho^0$ data in Ref.~\cite{Sibidanov:2013rkk} is shown multiplied by 
the factor $2\tau_0/\tau_+$.}
\label{tab:ed}
\end{table}

The theoretical input  consists of form factors values.
For $q^2$ in the  $[0,10]\,$GeV$^2$ range we will use the LCSR 
form factor values  obtained from the 
parameterizations  given in Ref.~\cite{Ball:2004rg}. 
 For higher $q^2$ we will use the lattice results by the 
SPQcdR ~\cite{Abada:2002ie} and 
UKQCD~\cite{Bowler:2004zb} collaborations. All of them  are collected in
Table~\ref{tab:ff}. For the LCSR form factors, and   
following Ref.~\cite{delAmoSanchez:2010af}, we have
assumed a 10\% error at $q^2=0$ that increases linearly to 13\% at 
$q^2=14\,{\rm
GeV}^2$. SPQcdR errors include both systematic and statistical uncertainties while in the
case of UKQCD data both statistical and systematic errors are shown. The latter
are highly asymmetric. Following Ref.~\cite{Flynn:2008zr}, and in order to
perform the fit, 
 we put the UKQCD form factors values in the center of their systematic
 range and  we  use half that range as the systematic error.

\begin{table}
\begin{tabular}{lllll}\hline\hline
&$q^2\,[{\rm GeV}^2]$&\hspace*{.8cm}$V$&\hspace*{.8cm}$A_1$&\hspace*{.8cm}$A_2$\\\hline
{\rm LCSR}~\cite{Ball:2004rg}
&0&$0.324\pm0.032$&$0.240\pm 0.024$&$0.221\pm0.022$\\
&1&$0.343\pm 0.035$  &  $0.247 \pm 0.025$  &  $0.232 \pm  0.024$\\
&2&$0.364\pm0.038$&$0.254\pm0.026$ &$0.244\pm0.025$\\
& 3&       $0.387\pm 0.041$  &  $0.261 \pm 0.028$  &  $0.257 \pm  0.027$\\
&4&$0.412\pm0.045$&$0.269\pm 0.029$&$0.271\pm 0.029$\\
&5&       $0.440\pm 0.049$  &  $0.277 \pm 0.031$ &   $0.286 \pm  0.032$\\
&6&$0.471\pm0.053$&$0.286\pm 0.032$&$0.302\pm 0.034$\\
&7  &     $0.506 \pm0.058$  &  $0.295 \pm 0.034$  &  $0.320 \pm  0.037$\\
&8&$0.546\pm 0.064$&$0.305\pm 0.036$&$0.339\pm 0.040$\\
&9 &      $0.590\pm 0.070$  &  $0.316 \pm 0.038$  & $ 0.360  \pm 0.043$\\
&10&$0.641\pm 0.078$&$0.327\pm 0.040$&$0.384\pm 0.047$\vspace{.25cm}\\
{\rm SPQcdR}~\cite{Abada:2002ie}\hspace{1cm}
&10.69&$0.51\pm0.26$&$0.354\pm0.085$&$0.38\pm0.26$\\
&12.02&$0.61\pm0.28$&$0.384\pm0.087$&$0.49\pm0.30$\\
&13.35&$0.74\pm0.30$&$0.421\pm0.089$&$0.65\pm0.35$\\
&14.68&$0.93\pm0.31$&$0.465\pm0.092$&$0.93\pm0.41$\\
&16.01&$1.20\pm0.32$&$0.519\pm0.097$&$1.41\pm0.56$\\
&17.34&$1.61\pm0.33$&$0.588\pm0.108$&$2.39\pm1.23$\\
&18.67&$2.26\pm0.55$&$0.678\pm0.134$&$4.7\pm4.1$\vspace{.25cm}\\
{\rm UKQCD}~\cite{Bowler:2004zb}
&$12.67$\hspace*{1.5cm}&$0.684\pm0.162^{+0.00}_{-0.56}$\hspace*{1cm}&$0.439\pm0.067^{+0.000}_{-0.080}$\hspace*{1cm}&$0.70\pm0.49^{+0.08}_{-0.03}$\\
&$13.01$&$0.714\pm0.162^{+0.00}_{-0.50}$&$0.448\pm0.065^{+0.000}_{-0.079}$&$0.71\pm0.46^{+0.08}_{-0.03}$\\
&$13.51$&$0.763\pm0.155^{+0.00}_{-0.40}$&$0.460\pm0.063^{+0.000}_{-0.075}$&$0.72\pm0.43^{+0.10}_{-0.02}$\\
&$14.02$&$0.818\pm0.147^{+0.00}_{-0.31}$&$0.472\pm0.059^{+0.000}_{-0.073}$&$0.73\pm0.42^{+0.12}_{-0.01}$\\
&$14.52$&$0.883\pm0.141^{+0.00}_{-0.24}$&$0.485\pm0.055^{+0.000}_{-0.070}$&$0.76\pm0.42^{+0.14}_{-0.03}$\\
&$15.03$&$0.967\pm0.137^{+0.00}_{-0.20}$&$0.498\pm0.051^{+0.000}_{-0.068}$&$0.78\pm0.46^{+0.16}_{-0.05}$\\
&$15.53$&$1.057\pm0.134^{+0.00}_{-0.19}$&$0.513\pm0.049^{+0.000}_{-0.067}$&$0.81\pm0.54^{+0.18}_{-0.06}$\\
&$16.04$&$1.164\pm0.150^{+0.10}_{-0.21}$&$0.529\pm0.047^{+0.000}_{-0.066}$&$0.84\pm0.71^{+0.20}_{-0.07}$\\
&$16.54$&$1.296\pm0.184^{+0.21}_{-0.25}$&$0.544\pm0.043^{+0.000}_{-0.062}$&$0.87\pm0.97^{+0.23}_{-0.08}$\\
&$17.05$&$1.46\pm0.26^{+0.34}_{-0.30}$ &$0.560\pm0.043^{+0.000}_{-0.059}$&$0.90\pm1.35^{+0.27}_{-0.07}$\\
&$17.55$&$1.67\pm0.40^{+0.49}_{-0.36}$&$0.577\pm0.043^{+0.000}_{-0.058}$&$0.90\pm1.89^{+0.33}_{-0.03}$\\
&$16.54$&$2.02\pm0.68^{+0.73}_{-0.48}$&$0.599\pm0.052^{+0.000}_{-0.058}$&$0.9\pm2.9^{+0.4}_{-0.1}$\\\hline\hline
\end{tabular}
\caption{Theoretical form factor inputs used for the fit. SPQcdR and UKQCD data
taken from Table 2 in Ref.~\cite{Flynn:2008zr}. Concerning UKQCD data see text for details.}
\label{tab:ff}
\end{table}
\subsection{$\chi^2$ definition}
The $\chi^2$ function we use for the fit is
\bea
\chi^2=\sum_{j,k=1}^{115}[(Q^{\rm input}_j-Q^{\rm fit}_j)C^{-1}_{jk}(Q^{\rm input}_k-Q^{\rm fit}_k)]
\eea
where  $Q^{\rm input}_j$ represents any of the input quantities and $Q^{\rm fit}_j$ 
is the corresponding value obtained in our
calculation. In order to construct the $C$ covariant matrix we have 
not considered any correlation between data from different experiments 
or between different theoretical calculations, or between experimental 
and theoretical inputs. $C$ is then block
diagonal. CLEO and BaBar collaborations provide  statistical and
systematic correlation matrices and in these two cases their corresponding 
blocks in
$C$ are constructed as
\bea
C_{jk}=\sigma_j\sigma_k{\cal C}^{\rm stat}_{jk}+\epsilon_j\epsilon_k{\cal
C}^{\rm sys}_{jk}.
\eea
with ${\cal C}^{{\rm stat}/{\rm sys}}$ the statistical/systematic 
correlation matrices.
The  Belle Collaboration~\cite{Sibidanov:2013rkk} also provides  two independent
 statistical
correlation matrices, one for $\rho^+$ data and one for $\rho^0$ data, so that we build
two independent blocks as
\bea
C_{jk}=\sigma_j\sigma_k{\cal C}^{\rm stat}_{jk}.
\eea
For the
block corresponding to UKQCD data we use
\bea
C_{jk}=\sigma^2_j\delta_{jk}+\epsilon_j\epsilon_k,
\eea 
that assumes independent statistical uncertainties and fully correlated
systematic errors.
Finally for LCSR and SPQcdR results we
use
\bea
C_{jk}=\sigma^2_j\delta_{jk}.
\eea
\section{Results and discussion}
\label{sec:res}
Best fit results are compiled in Table~\ref{tab:res}.
\begin{table}
\begin{tabular}{ll}\hline\hline
$|V_{ub}|$&$(3.12\pm0.13)\times 10^{-3}$\\
$V(0)$&$0.349\pm0.022$\\
$V(2q^2_{\rm max}/3)$\hspace*{.5cm}&$0.863\pm0.042$\\
$V(q^2_{\rm max})$& $2.553\pm0.375$\\
$A_1(0)$&$0.253\pm 0.011$\\
$A_1(2q^2_{\rm max}/3)$&$0.429\pm 0.013$\\
$A_1(q^2_{\rm max})$&$0.719\pm 0.038$\\
$A_2(0)$&$0.226\pm 0.014$\\
$A_2(2q^2_{\rm max}/3)$&$0.701\pm 0.065$\\
$A_2(q^2_{\rm max})$&$3.033\pm0.914$\\\hline\hline
\end{tabular}
\caption{Best fit parameters of the global fit. }\vspace{.5cm}
\label{tab:res}
\end{table}
 \begin{figure}[h!!!!!]
\begin{center}
\resizebox{6.5cm}{!}{\includegraphics{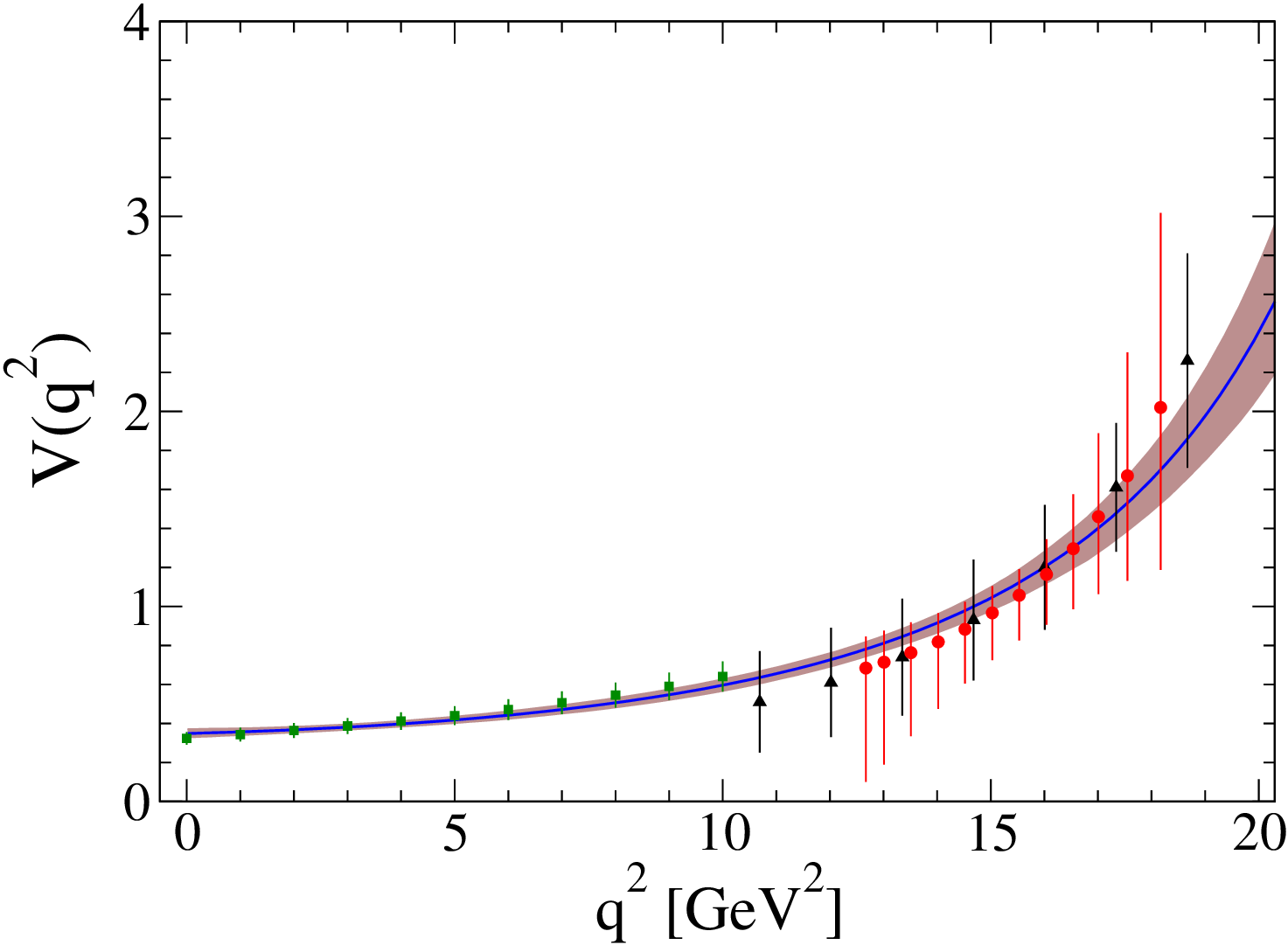}}\hspace{.5cm}
\resizebox{6.5cm}{!}{\includegraphics{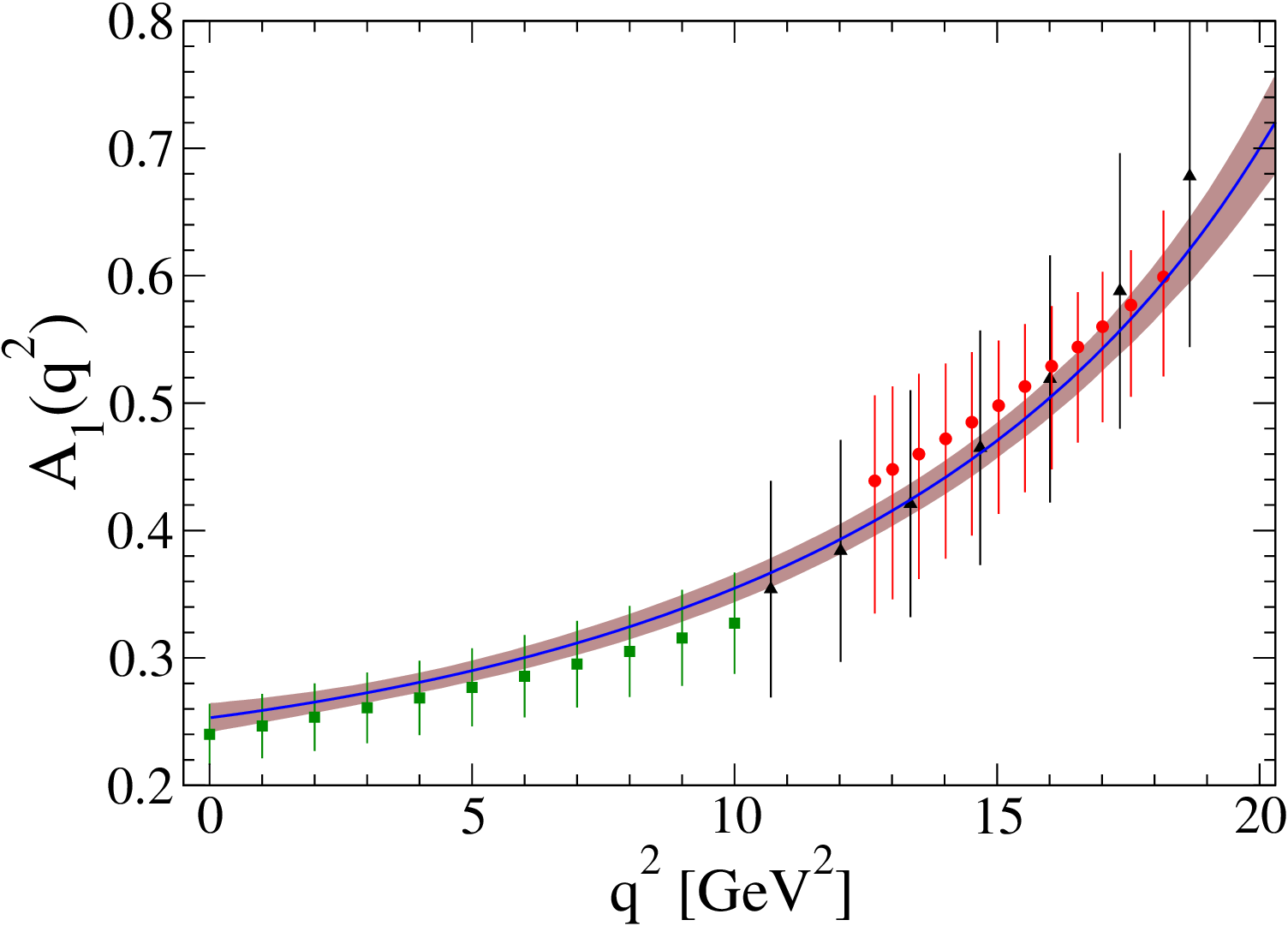}}\vspace{1cm}
{\resizebox{6.5cm}{!}{\includegraphics{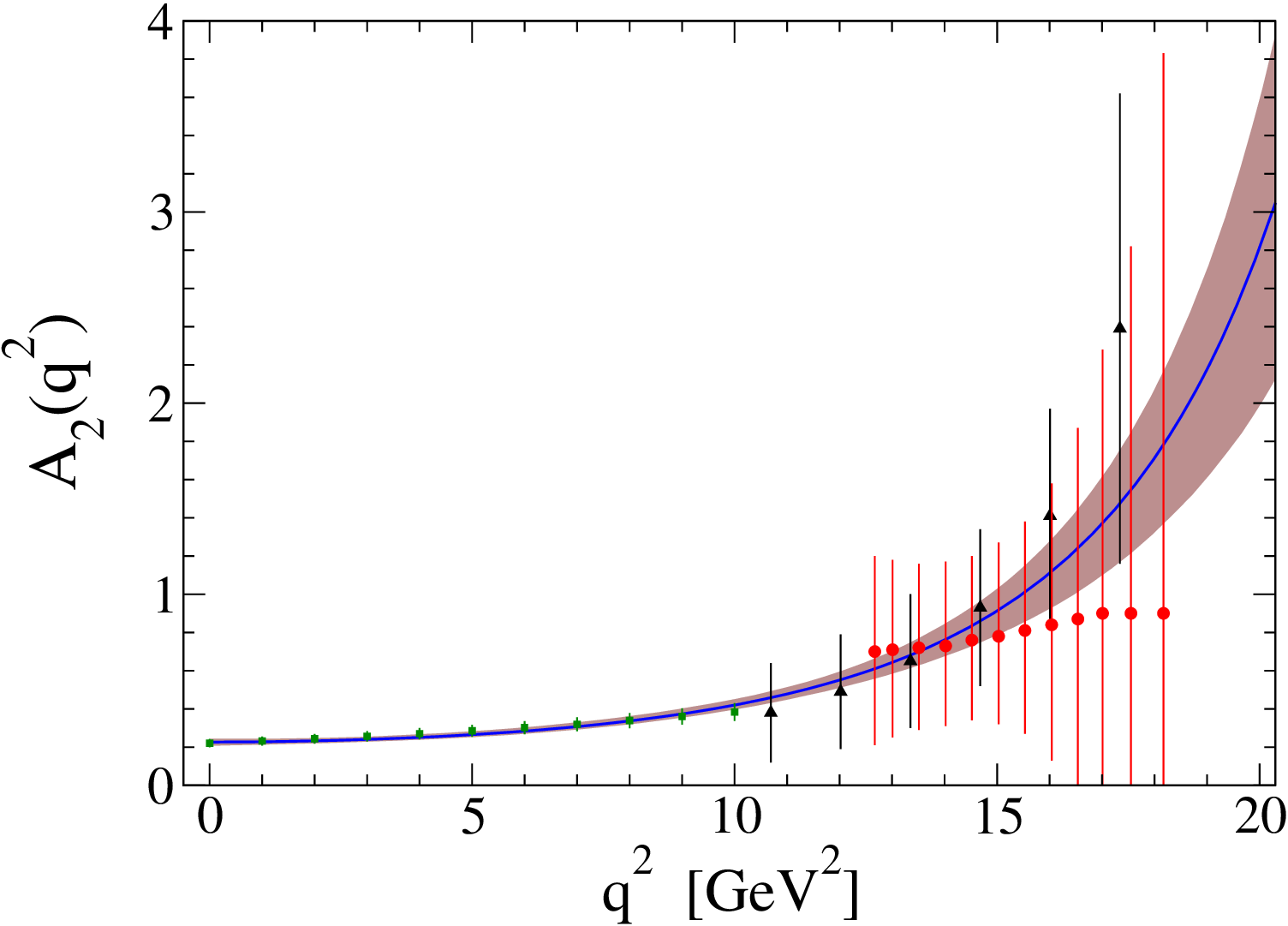}}}\hspace{.5cm}
\resizebox{6.5cm}{!}{\includegraphics{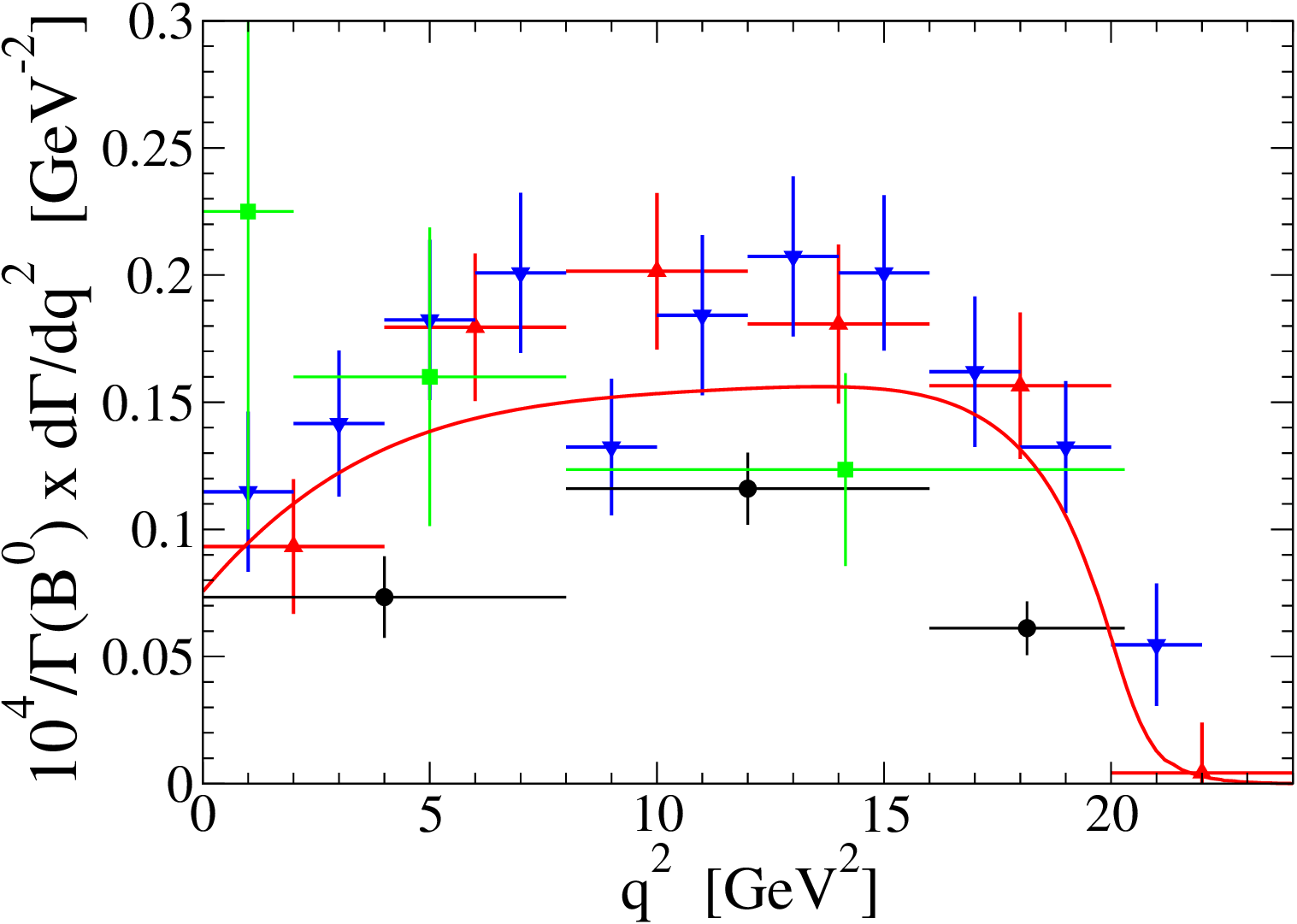}}
\vspace{1cm}\\
\caption{Top panels and left-bottom panel: Form factor  obtained from
 the fit (solid  line) together with their corresponding 68\%
confidence level band. We show the predictions from LCSR~\cite{Ball:2004rg}
(squares), and lattice QCD from the SPQcdR~\cite{Abada:2002ie}
 (up-triangles) and UKQCD~\cite{Bowler:2004zb} (circles)
collaborations. Right-bottom panel: $\frac{10^4}{\Gamma(B^0)}\frac{d\Gamma}{dq^2} 
 [{\rm GeV}^{-2}]$. 
 Up-triangles, down-triangles, 
circles,
 and squares stand respectively for Belle $\rho^+$ and $\rho^0$ data~\cite{Sibidanov:2013rkk}, BaBar data~\cite{delAmoSanchez:2010af}, 
 and CLEO data~\cite{Gray:2007pw}. The solid line stands for our prediction.}
\label{fig:ffg}
\end{center}
\end{figure} 
The fit has $\chi^2/{\rm d.o.f.}=1.6$ for a total of 105 degrees of freedom.
The corresponding Gaussian correlation matrix is given in
Appendix~\ref{app:gcm}.
The value $|V_{ub}|=(3.12\pm0.13)\times 10^{-3}$ extracted from our
global fit analysis is in
agreement with the average determination from the exclusive $B\to\pi$
 decays given by $(3.23\pm0.31)\times
10^{-3}$~\cite{Beringer:1900zz}. A calculation based in
Eq.(\ref{eq:rhosinanchura}), i.e. ignoring $\rho$ meson width effects,
 would have provided a smaller value of $|V_{ub}|=(2.87\pm0.13)\times 10^{-3}$.

In Fig.~\ref{fig:ffg} 
we
show the form factors, together with their 68\% confidence
level bands, that result from the global fit,  and we compare them  to the different
theoretical input. Finally in Fig.~\ref{fig:ffg}
(bottom-right panel)
we also present our prediction for $\frac{10^4}{\Gamma(B^0)}\frac{d\Gamma}{dq^2}$ and
compare it to data by the Belle~\cite{Sibidanov:2013rkk}, BaBar~\cite{delAmoSanchez:2010af}, 
 and CLEO~\cite{Gray:2007pw} collaborations.  The largest
discrepancy occurs for CLEO data where the experimental 
 distribution peaks a significantly smaller $q^2$ values
  than the theoretical distribution. This seems to be incompatible with 
  the theoretical form factor 
 predictions at low $q^2$ obtained in LCSR. 
 Belle and BaBar results agree better in shape with our analysis. However,
  one clearly sees
  in Fig.~\ref{fig:ffg} that BaBar data would prefer a smaller $|V_{ub}|$ value, whereas
  Belle data would be better reproduced with a higher $|V_{ub}|$
  value.

%
The recent data by the Belle Collaboration~\cite{Sibidanov:2013rkk} 
 gives results for
 smaller $q^2$ bins which means more data and then the possibility for 
 more stringent
 constraints on theoretical models. In this respect it is worth
 making a fit just to Belle's data together with the form factors. 
In this case one gets   $|V_{ub}|=(3.51\pm0.16)\times 10^{-3}$ 
which  is in perfect
agreement with the analyses in Ref.~\cite{Sibidanov:2013rkk} where other
sets of form factors were used. 
  A fit to the form factors and to the BaBar
 data of
 Ref.~\cite{delAmoSanchez:2010af}  alone
 would give $|V_{ub}| = (2.52\pm 0.18)\times 10^{-3}$ which is much smaller than
 the result obtained from Belle data.  Note also that the total decay rate from 
 BaBar extracted in Ref.~\cite{delAmoSanchez:2010af} is  some 15\% smaller 
 that the one provided in their earlier 
 measurement of   Ref.~\cite{Aubert:2005cd} and used in \cite{Flynn:2008zr}.

\begin{figure}[h!!]
\begin{center}
\vspace{1cm}\resizebox{10.cm}{!}{\includegraphics{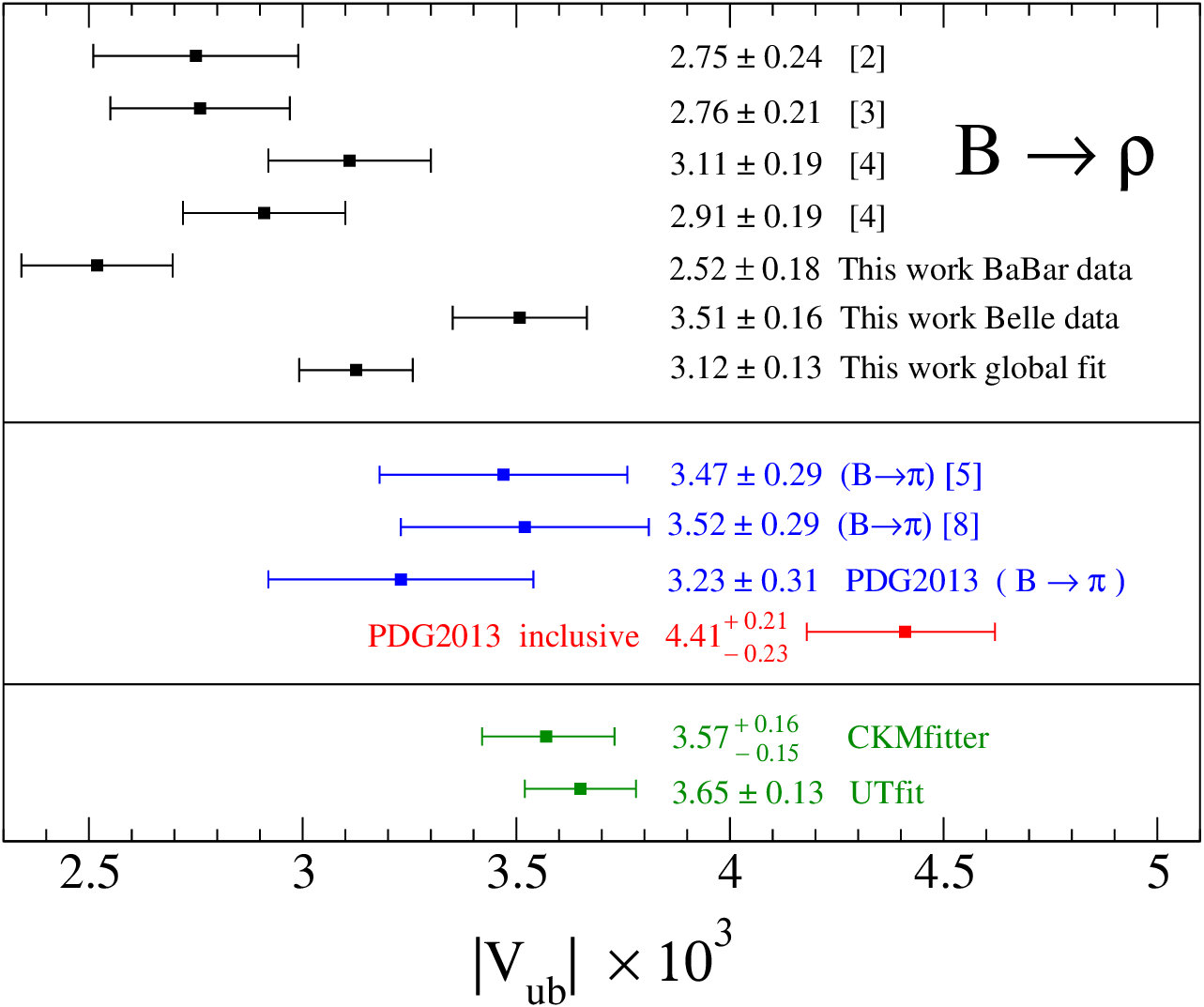}}
\caption{Different $|V_{ub}|$ values obtained in $B\to\rho$ decay analyses. We
also show for comparison the $|V_{ub}|$ determination from the $B\to\pi$ decay
in Refs.~\cite{Flynn:2007ii,Sibidanov:2013rkk}, the PDG exclusive and 
inclusive 2013 average 
 updates~\cite{Beringer:1900zz}, and the fits from the
 CKMfitter~\cite{ckmfitter} and UTfit~\cite{utfit} groups. }
\label{fig:vub}
\end{center}
\end{figure}

In Fig.~\ref{fig:vub} we show different $|V_{ub}|$ values obtained 
in $B\to\rho$ decay analyses.  As mentioned, our global fit result is  in  agreement
with the  $B\to\pi$ exclusive decay 
average value quoted in the PDG 2013 
update~\cite{Beringer:1900zz}. The results by the CKMfitter~\cite{ckmfitter} and UTfit~\cite{utfit} 
groups are in very good agreement with our determination using only
the recent Belle data.
However, as seen in Fig.~\ref{fig:vub}, there still persist the large
discrepancy between inclusive and exclusive determinations of $|V_{ub}|$, with
the
global fits by the CKMfitter~\cite{ckmfitter} and UTfit~\cite{utfit}  groups
being in better agreement with the latter.

\begin{acknowledgments}
 This research was supported by  the Spanish Ministerio de Econom\'{\i}a y 
 Competitividad and European FEDER funds
under Contracts Nos. FPA2010-21750-C02-02,  FIS2011-28853-C02-02,  
and the Spanish Consolider-Ingenio 2010 Programme CPAN (CSD2007-00042), by Generalitat
Valenciana under Contract No. PROMETEO/20090090, by Junta de Andalucia under
Contract No. FQM-225,
 by the EU HadronPhysics3 project, Grant Agreement
No. 283286,  and by the
University of Granada start-up Project for Young Researches contract No. PYR-2014-1.
C.A. wishes to acknowledge a CPAN postdoctoral contract.
  \end{acknowledgments}

%
%
\appendix
\section{Helicity amplitudes}
\label{app:ha}
In this appendix we shall write the product 
${\cal L}^{\alpha\alpha'}(\tilde p_l, 
\tilde p_\nu)
{\cal H}_{\alpha\alpha'}(\Lambda p_B,\Lambda p_\rho )$ in terms of helicity
amplitudes. For that purpose we use that
\bea
g^{\mu\nu}=\sum_{r=t,\pm1,0}g_{rr}\,\epsilon_r^\mu(\Lambda q)
\epsilon_r^{\nu\, *}(\Lambda q)=\sum_{r=t,\pm1,0}g_{rr}\,\epsilon_r^{\mu\, *}(\Lambda q)
\epsilon_r^{\nu}(\Lambda q),
\eea
with $g_{tt}=-g_{+1+1}=-g_{-1-1}=-g_{00}=1$ and
\bea
\epsilon_t(\Lambda q)&=&\frac{\Lambda q}{\sqrt{q^2}}=(1,0,0,0),\\
\epsilon_{+1}(\Lambda q)&=&\Big(0,\frac{-1}{\sqrt2},\frac{-i}{\sqrt2},0\Big),\\
\epsilon_{-1}(\Lambda q)&=&\Big(0,\frac{1}{\sqrt2},\frac{-i}{\sqrt2},0\Big),\\
\epsilon_{0}(\Lambda q)&=&(0,0,0,1).
\eea
Then,
\bea
{\cal L}^{\alpha\alpha'}(\tilde  p_l,\tilde  p_\nu)
\,{\cal H}_{\alpha\alpha'}(\Lambda p_B,
\Lambda p_\rho )=\sum_{r=t,\pm1,0}\sum_{s=t,\pm1,0} g_{rr}g_{ss}\,
{\cal L}_{rs}(\tilde p_l,\tilde p_\nu)\,
{\cal H}_{rs}(\Lambda p_B,
\Lambda p_\rho ),
\eea
where we have defined the hadronic and leptonic helicity amplitudes
\bea
{\cal H}_{rs}(\Lambda p_B,
\Lambda p_\rho )&=&\epsilon_r^{\alpha\, *}(\Lambda q)\,
{\cal H}_{\alpha\alpha'}(\Lambda p_B,
\Lambda p_\rho )\,\epsilon_s^{\alpha'}(\Lambda q),\\
{\cal L}_{rs}(\tilde  p_l,\tilde p_\nu )&=&\epsilon_r^{\beta}(\Lambda q)\,
{\cal L}_{\beta\beta'}(\tilde  p_l,\tilde  p_\nu )\,\epsilon_s^{\beta'\, *}(\Lambda q).
\eea
As
\bea
{\cal H}_{\alpha\alpha'}(\Lambda p_B,\Lambda p_\rho)=
\sum_{u=\pm1,0}h_{\alpha\sigma}(\Lambda p_B,\Lambda
p_\rho)\epsilon_{u}^{\sigma\,*}(\Lambda p_\rho)
h_{\alpha'\sigma'}^*(\Lambda p_B,\Lambda
p_\rho)\epsilon_{u}^{\sigma'}(\Lambda p_\rho),
\eea
we will have
\bea
{\cal H}_{rs}(\Lambda p_B,
\Lambda p_\rho )&=&\sum_{u=\pm1,0}\epsilon_r^{\alpha\,*}(\Lambda q)h_{\alpha\sigma}(\Lambda p_B,\Lambda
p_\rho)\epsilon_{u}^{\sigma\,*}(\Lambda p_\rho)
\epsilon_s^{\alpha'}(\Lambda q)
h_{\alpha'\sigma'}^*(\Lambda p_B,\Lambda p_\rho)\epsilon_{u}^{\sigma'}(\Lambda p_\rho)\\
&=&\sum_{u=\pm1,0}h_{ru}(\Lambda p_B,\Lambda p_\rho)
h_{su}^*(\Lambda p_B,\Lambda p_\rho),
\eea
with
\bea
h_{ru}(\Lambda p_B,\Lambda p_\rho)=\epsilon_r^{\alpha\,*}(\Lambda q)h_{\alpha\sigma}(\Lambda p_B,\Lambda
p_\rho)\epsilon_{u}^{\sigma\,*}(\Lambda p_\rho).
\eea
Using
\bea
\epsilon_{+1}(\Lambda p_\rho)&=&\Big(0,\frac{-1}{\sqrt2},\frac{-i}{\sqrt2},0\Big),\\
\epsilon_{-1}(\Lambda p_\rho)&=&\Big(0,\frac{1}{\sqrt2},\frac{-i}{\sqrt2},0\Big),\\
\epsilon_{0}(\Lambda
p_\rho)&=&\Big(\frac{\lambda^{1/2}(q^2,m_B^2,p_\rho^2)}{2\sqrt{q^2}\sqrt{p_\rho^2}}
,0,0,-\frac{m_B^2-q^2-p_\rho^2}{2\sqrt{q^2}\sqrt{p_\rho^2}}\Big).
\eea
we can evaluate the $h_{ru}$ quantities. The nonzero ones are
\bea
h_{t0}&=&\frac{\lambda^{1/2}(m_B^2,q^2,p_\rho^2)}{2\sqrt{p_\rho^2}\sqrt{q^2}}
\Big[-iA_1(q^2)\Big(m_B+\sqrt{p_\rho^2}\,\Big)+iA_2(q^2)
\Big(m_B-\sqrt{p_\rho^2}\ \Big)
-i2A(q^2)\sqrt{p_\rho^2}\ \Big],\nonumber\\
h_{+1-1}&=&-iA_1(q^2)\Big(m_B+\sqrt{p_\rho^2}\,\Big)
+iV(q^2)\frac{\lambda^{1/2}(m_B^2,q^2,p_\rho^2)}{m_B+\sqrt{p_\rho^2}}
,\nonumber\\
h_{-1+1}&=&-iA_1(q^2)\Big(m_B+\sqrt{p_\rho^2}\,\Big)
-iV(q^2)
\frac{\lambda^{1/2}(m_B^2,q^2,p_\rho^2)}{m_B+\sqrt{p_\rho^2}},\nonumber\\
h_{00}&=&\frac1{2\sqrt{p_\rho^2}\sqrt{q^2}}
\Big[-iA_1(q^2)\Big(m_B+\sqrt{p_\rho^2}\,\Big)(m_B^2-p_\rho^2-q^2)
+i\frac{A_2(q^2)\lambda(m_B^2,q^2,p_\rho^2)}{m_B+\sqrt{p_\rho^2}}\,
\Big].
\eea
From these values we get the following nonzero hadronic helicity amplitudes
\bea
{\cal H}_{tt}&=&\frac{\lambda(m_B^2,q^2,p_\rho^2)}{4\,{p_\rho^2}\,{q^2}}
\left[-A_1(q^2)\Big(m_B+\sqrt{p_\rho^2}\,\Big)
+A_2(q^2)\Big(m_B-\sqrt{p_\rho^2}\,\Big)
-2A(q^2)\sqrt{p_\rho^2}\ \right]^2,\nonumber\\
{\cal H}_{t0}={\cal H}_{0t}&=&\frac{\lambda^{1/2}(m_B^2,q^2,p_\rho^2)}
{4\,{p_\rho^2}\,{q^2}}
\Big[-A_1(q^2)\Big(m_B+\sqrt{p_\rho^2}\,\Big)(m_B^2-p_\rho^2-q^2)
+\frac{A_2(q^2)\lambda(m_B^2,q^2,p_\rho^2)}
{m_B+\sqrt{p_\rho^2}}\
\Big]\nonumber\\
&&\hspace{.5cm}\times\Big[-A_1(q^2)\Big(m_B+\sqrt{p_\rho^2}\,\Big)
+A_2(q^2)\Big(m_B-\sqrt{p_\rho^2}\,\Big)
-2A(q^2)\sqrt{p_\rho^2}\ \Big],\nonumber\\
{\cal H}_{00}&=&\frac1{4\,{p_\rho^2}\,{q^2}}
\Big[-A_1(q^2)\Big(m_B+\sqrt{p_\rho^2}\,\Big)(m_B^2-p_\rho^2-q^2)
+\frac{A_2(q^2)\lambda(m_B^2,q^2,p_\rho^2)}{m_B+\sqrt{p_\rho^2}}
\ \Big]^2,\nonumber\\
{\cal H}_{+1+1}&=&\Big[A_1(q^2)\Big(m_B+\sqrt{p_\rho^2}\,\Big)
-V(q^2)\frac{\lambda^{1/2}(m_B^2,q^2,p_\rho^2)}{m_B+\sqrt{p_\rho^2}}
\ \Big]^2,\nonumber\\
{\cal H}_{-1-1}&=&\Big[A_1(q^2)\Big(m_B+\sqrt{p_\rho^2}\,\Big)
+V(q^2)\frac{\lambda^{1/2}(m_B^2,q^2,p_\rho^2)}{m_B+\sqrt{p_\rho^2}}
\ \Big]^2.
\eea

The corresponding leptonic helicity amplitudes are given by
\bea
{\cal L}_{tt}&=&(q^2-m_l^2)\frac{m_l^2}{2\,{q^2}},\nonumber\\
{\cal L}_{t0}={\cal L}_{0t}&=&-(q^2-m_l^2)x_l\frac{m_l^2}
{2\,{q^2}},\nonumber\\
{\cal L}_{00}&=&(q^2-m_l^2)\frac1{2}\Big[1-x_l^2+x_l^2\frac{m_l^2}{q^2}\Big],\nonumber\\
{\cal L}_{+1+1}&=&(q^2-m_l^2)\frac1{4}\Big[(1\mp x_l)^2+
(1-x_l^2)\,\frac{m_l^2}{q^2}
\Big],\nonumber\\
{\cal L}_{-1-1}&=&(q^2-m_l^2)\frac1{4}\Big[(1\pm x_l)^2+
(1-x_l^2)\,\frac{m_l^2}{q^2}
\Big],
\eea
where the upper (lower) sign corresponds to $l^+\nu_l$ ($l^-,\bar\nu_l$ ) decays. 
\section{Gaussian correlation matrix}
\label{app:gcm}
The Gaussian correlation matrix corresponding to the best fit parameters in
Table~\ref{tab:res} reads
\bea
\left(\begin{array}{rrrrrrrrrr}
1.0000&-0.0027 & -0.3447 &    -0.2775 &    -0.1698   &  -0.6730  &   -0.2685 
&0.1366 &    0.3584 &    0.2873 \\
&1.0000&-0.2204  &   0.3410 &    -0.0483 &0.0250& -0.0649
&0.0112& -0.0087& -0.0029\\
&&1.0000&0.4401  &   0.0627& 0.3049 &    0.2435  &   -0.0199& 0.1439 &    0.1555  \\
&&&1.0000 &-0.0105& 0.3178 &    0.0670& -0.0002& 0.1273 &    0.1544 \\
&&&&1.0000 &-0.1705 &    0.4191  &   0.3714   &  -0.2169&     -0.1325 \\
&&&&&1.0000 &0.2644  &   -0.1534  &   0.1291   &  0.0660\\
&&&&&&1.0000 &0.1610 &    0.1796  &   0.3333 \\
&&&&&&&1.0000 & 0.0989& 0.3782 \\
&&&&&&&&1.0000 & 0.8677  \\
&&&&&&&&& 1.0000\\
\end{array}
\right)
\eea

%
%
%
%
%
%
%
%
%

%
%
%
%

%
%
%
%

\end{document}